\documentclass[pre,twocolumn,amsmath,showpacs,superscriptaddress]{revtex4}

\usepackage{color} 

\usepackage{amsmath}
\usepackage{amsbsy}
\usepackage{amscd}
\usepackage{amsopn}
\usepackage{amstext}
\usepackage{amsxtra}
\usepackage{epsfig}

\newcommand{\WT}{\widetilde}

\def\kp{{k^\prime}}

\def\lesssim{\mathrel{\hbox{\rlap{\hbox{\lower4pt\hbox{$\sim$}}}\hbox{$<$}}}}

\def\g{\gamma}

\begin{document}

\title{Accuracy of Mean-Field Theory for Dynamics on Real-World Networks}

\author{James P. Gleeson}
\affiliation{MACSI, Department of Mathematics \& Statistics, University of Limerick, Ireland}
\author{Sergey Melnik}
\affiliation{MACSI, Department of Mathematics \& Statistics, University
of Limerick, Ireland}
\affiliation{Oxford Centre for Industrial and Applied Mathematics, Mathematical Institute, University of Oxford, OX1 3LB, UK}
\affiliation{CABDyN Complexity Centre, University of Oxford, Oxford, OX1 1HP, UK}
\author{Jonathan A. Ward}
\affiliation{MACSI, Department of Mathematics \& Statistics,
University of Limerick, Ireland}
\affiliation{Centre for Mathematics of Human Behaviour, Department of Mathematics \& Statistics, University of Reading, Reading, RG6 6AX, UK}
\author{Mason A. Porter}
\affiliation{Oxford Centre for Industrial and Applied Mathematics, Mathematical Institute, University of Oxford, OX1 3LB, UK}
\affiliation{CABDyN Complexity Centre, University of Oxford, Oxford, OX1 1HP, UK}
\author{Peter J. Mucha}
\affiliation{Carolina Center for Interdisciplinary Applied Mathematics, Department of Mathematics, University of North Carolina, Chapel Hill, NC 27599-3250, USA}
\affiliation{Institute for Advanced Materials, Nanoscience \& Technology, University of North Carolina, Chapel Hill, NC 27599-3216, USA}


\newcommand{\pjm}[1]{{\color{red}\emph{pjm: #1}}}


\date{19 Dec 2011} 



\begin{abstract}

Mean-field analysis is an important tool for understanding dynamics on complex networks. However, surprisingly little attention has been paid to the question of whether mean-field predictions are accurate, and this is particularly true for real-world networks with clustering and modular structure.  In this paper, we compare mean-field predictions to numerical simulation results for dynamical processes running on 21 real-world networks and  demonstrate  that the accuracy of the theory depends not only on the mean degree of the networks but also on the mean first-neighbor degree.  We show that mean-field theory can give (unexpectedly) accurate results for certain dynamics on disassortative real-world networks even when the mean degree is as low as 4.

\end{abstract}

\maketitle


\section{Introduction}

Mean-field theories are the most common form of analytical
approximation employed when studying dynamics on complex networks \cite{Barrat08}.
Typically, mean-field theories are derived under several
assumptions:
\begin{enumerate}
\item[(i)] {\emph{Absence of local clustering.}} When considering possible changes to the state of a
node $B_1$, it is assumed that the states of the
neighbors of node $B_1$ are independent of each other.  However, this assumption that the network is ``locally
tree-like'' does not hold if the neighbors of $B_1$
are also linked to each other---i.e., if the network is clustered (exhibits transitivity).
\item[(ii)] {\emph{Absence of modularity.}} It is also usually assumed that all nodes of the same degree $k$ are well-described
by the mean $k$-class state---i.e. by the average over all nodes of
degree $k$. However, this might not be true if the network is modular, so
that the states of degree-$k$ nodes are differently distributed
in different communities.
\item[(iii)] {\emph{Absence of dynamical correlations.}} Finally, it is assumed that the states of each node $B_1$ and those of its neighbors can be treated as independent when updating the state of node $B_1$.
\end{enumerate}
Importantly, the neglect of {\emph{dynamical}} correlations (as distinct from
{\emph{structural}} correlations such as degree-degree correlations \cite{Newman02,Estrada11}) between
neighbors in assumption (iii) can be addressed in improved theories that incorporate
information on the joint distribution of node \emph{states} at the ends
of a random edge in the network \cite{Newmanbook,Durrett10} (cf. theories that only specify the \emph{structures} at the ends of a random edge). The improved theories are often called \emph{pair-approximations}
(PA)  (examples are \cite{Vazquez08,Pugliese09}), and these are inevitably more complicated to derive and study than mean-field (MF) theories, so we mostly restrict our attention in the present paper to the more common MF-theory situation \footnote{One should not think of PA theories as theories that only add information about degree-degree correlations. Each particular PA or MF theory can either neglect degree-degree correlations (or other \emph{structural} correlations, that specify connections between nodes), or take them into account. The difference is that MF theories are derived under assumption (iii) of the absence of \emph{dynamical} correlations while PA theories take the dynamical correlations into consideration.}.


The distinctions between assumptions (i)---(iii) can be clarified by considering the theoretical approaches beyond the MF level that have been developed in certain cases to deal with violations of (i), (ii), and (iii). The impact of non-zero clustering on percolation problems on a network has been examined in Refs.~\cite{Newman09, Miller09, Gleeson09a, Gleeson10, Karrer10}.  The analytical methods used in those papers explicitly account for the dependence of neighbors' states on each other---i.e., for the violation of MF assumption (i). The role of community or modular structures [see assumption (ii)] on percolation requires a different extension of analytical methods \cite{Newman03, Gleeson08a}. As noted above, PA theories can account for dynamical correlations more accurately than MF theory, thereby improving on assumption (iii).

The MF assumptions enumerated above are clearly violated for
real-world networks, which are often highly clustered and modular
\cite{Newmanbook,comnotices}.  It is therefore rather surprising that MF theory
often provides a reasonably good approximation to the actual dynamics
on many real-world networks.  This fact has been noted by several
authors \cite{Barrat08,Dorogovtsev08,Durrett10,baron06,baron10}, but to our knowledge
no comprehensive explanation for this phenomenon has ever been
developed. In studying this phenomenon, we focus on a specific question of obvious
practical interest: Given a real-world network and a dynamical process
running on it, is it possible to predict whether or not MF theory will
provide a good approximation to the actual dynamics on this network?
In this paper, we test multiple well-studied dynamical processes on 21
undirected, unweighted real-world networks.  We enumerate and summarize various properties of the networks in Table~\ref{table1}.  These networks are characterized by a range of values for several standard network diagnostics, which is important for this study.
We show that MF theory typically works well provided $d$, the mean
degree of first neighbors of a random node, is sufficiently large.  In
contrast, we demonstrate that the mean degree $z$ of the network is
not necessarily a good indicator of MF accuracy.

The remainder of this paper is organized as follows. In
Section~\ref{sec2}, we introduce the dynamical processes that we
consider and compare numerical results with MF theory for sample
real-world networks. In Section~\ref{sec3}, we discuss the
implications of our results and propose an explanatory hypothesis.  In
Section~\ref{sec4}, we compare our results with earlier work in this
area. We conclude in Section~\ref{sec5}.


\section{Examples} \label{sec2}

We begin by showing examples for which MF theory gives accurate results for dynamics on real-world networks, contrasting with examples in which MF theory performs poorly.


\subsection{Kuramoto Phase Oscillator Model}

\begin{figure}
\centering \epsfig{figure=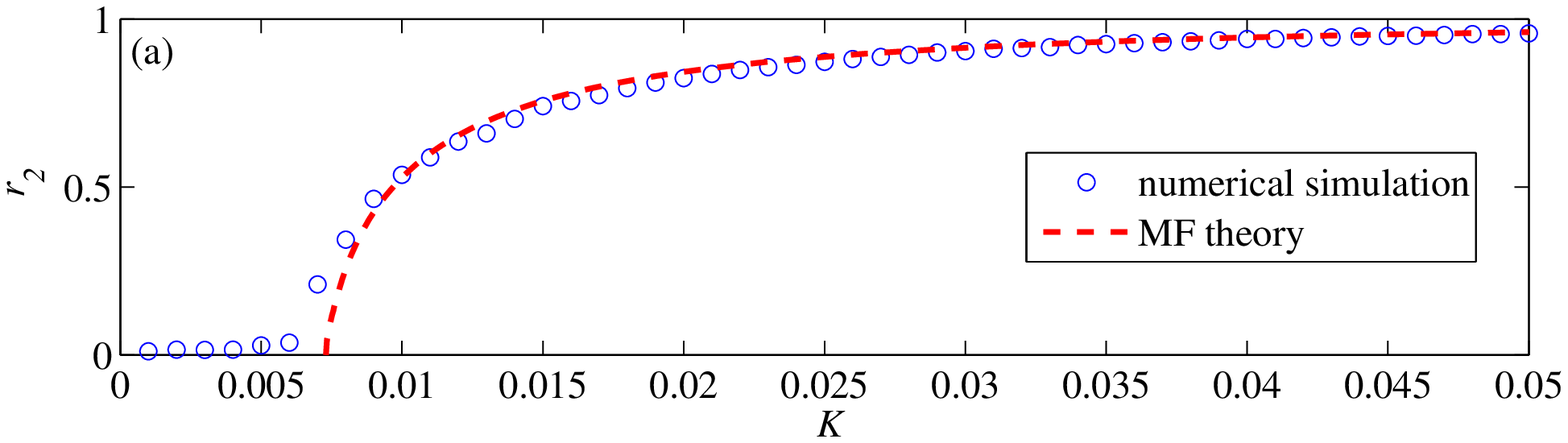,width=9.2cm}
\epsfig{figure=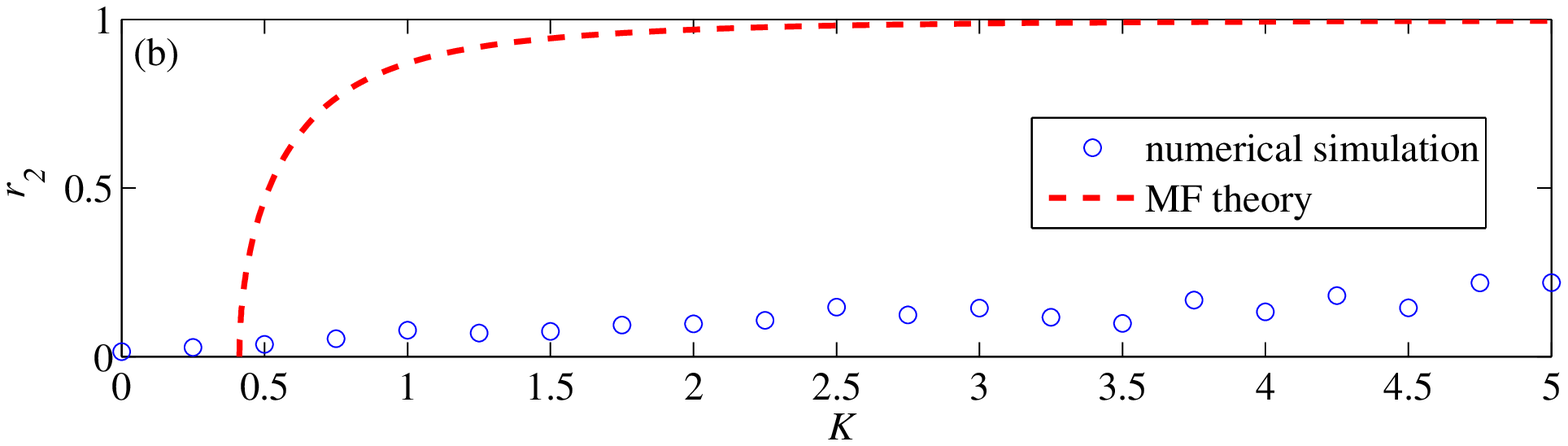,width=9.2cm}
 \caption{(Color online) Order parameter for synchronization in the Kuramoto phase oscillator
 model running on (a) the Facebook Oklahoma network \cite{Traud08} and (b) the US power grid network \cite{Watts98} as a function of the coupling $K$. The order parameter $r_2$ is defined in Eq.~(\ref{r2eqn}). The MF theory of \cite{Ichinomiya04} is given by Eq.~(\ref{eqn1}).} \label{fig1}
\end{figure}

In Fig.~\ref{fig1}, we show the results of running the Kuramoto phase
oscillator model \cite{Kuramotoref} on the Facebook Oklahoma network \cite{Traud08} and on the US
Power Grid network \cite{Watts98}. Each node corresponds to an oscillator with an intrinsic frequency drawn from a unit-variance Gaussian distribution. The phase $\theta_j(t)$ of the oscillator at node $j$ obeys the differential equation
 \begin{equation}
	 \frac{d \theta_j}{d t} = \omega_j+ K \sum_{\ell=1}^N A_{j \ell} \sin\left(\theta_\ell-\theta_j\right)\,,
 \end{equation}
 where $\omega_j$ is the intrinsic frequency of node $j$, $N$ is the number of nodes, and $A$ is the adjacency matrix of the network.  The coupling to network neighbors is measured by the
parameter $K$, and global synchrony of
the oscillators is expected to emerge for sufficiently large $K$ \cite{arenas08}. Synchrony is quantified using the order parameter \cite{Ichinomiya04}
 \begin{equation}
	r_2 = \frac{\left|\sum_{j=1}^N k_j e^{i \theta_j}\right|}{\sum_{j=1}^N k_j}\,,\label{r2eqn}
 \end{equation}
where $k_j$ is the degree of node $j$ and $i=\sqrt{-1}$.  The MF theory of Ref.~\cite{Ichinomiya04} (see also \cite{Restrepo05}) yields the following implicit equation for $r_2$:
\begin{align}
	\sum_{k=0}^\infty k^2 p_k e^{-K^2 k^2 r_2^2/4}\left[ I_0\left(\frac{K^2 k^2 r_2^2}{4}\right)+ I_1\left(\frac{K^2 k^2 r_2^2}{4}\right) \right]\nonumber\\
\text{ }&\hspace{-2.5cm}=\frac{2\sqrt{2}z}{\sqrt{\pi} K}\,, \label{eqn1}
\end{align}
where $I_n$ is the modified Bessel function of the first kind,  $p_k$ is the degree distribution of the network, and $z =\left<k\right>\equiv \sum k p_k$ is the mean degree.  The agreement in Fig.~\ref{fig1} between theory and simulation is very good for the Facebook network but very poor
for the Power Grid.  (See Fig.~\ref{fig4} for additional examples.)

The results of  Fig.~\ref{fig1} are perhaps explained in part by noting that the mean
degree $z$ of the Facebook network is 102, whereas $z \approx 2.67$ for the Power
Grid (see Table~\ref{table1}). It is arguable that nodes with many neighbors will experience something closer to a ``mean field'' than nodes with few neighbors.  In particular, it
is plausible that low-$z$ networks might be more prone to errors in MF due to neglecting the effects of clustering, modularity, and dynamical correlations. This is attractively simple, but as we show below, this naive explanation does not fully capture certain subtleties of this question.


\subsection{SIS Epidemic Model}

 \begin{figure}
\centering
\epsfig{figure=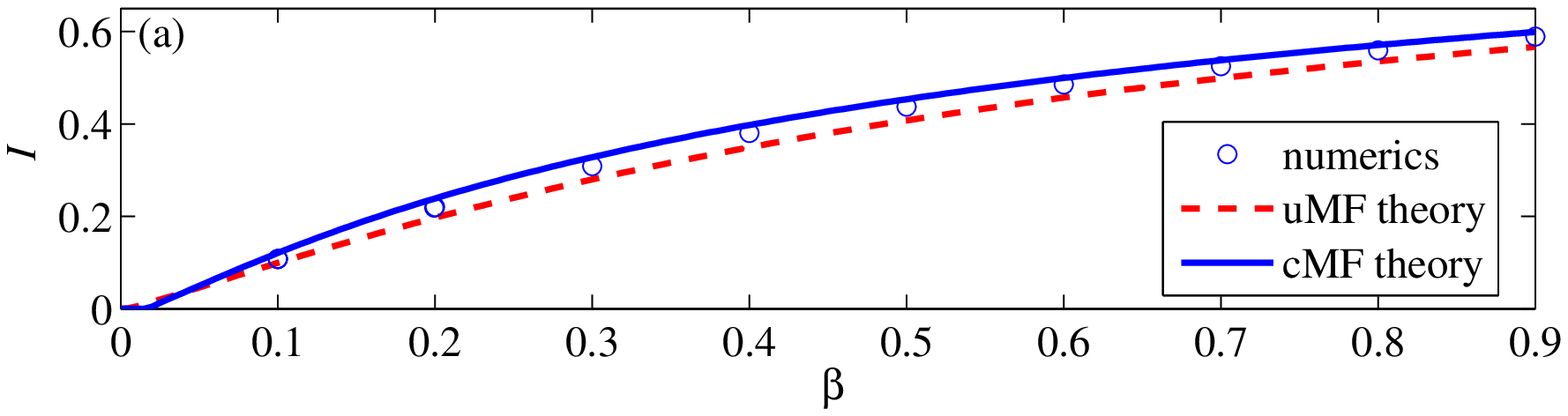,width=9.5cm}
\epsfig{figure=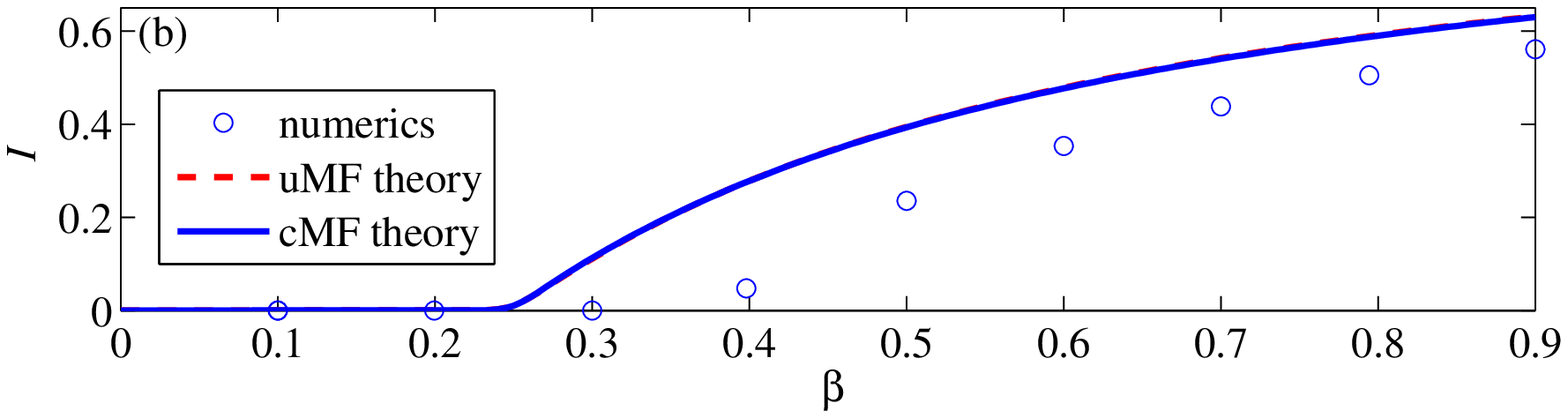,width=9.5cm}
 \caption{(Color online) Fraction of infected nodes in the steady state of the SIS process on the (a) AS Internet \cite{asrel_url} and (b) Electronic Circuit (s838) \cite{Milo04}
 networks as a function of the spreading rate $\beta$. The MF theory is from Refs.~\cite{PastorSatorras01,Barthelemy05}.  Observe that uncorrelated and correlated MF theories are indistinguishable in panel (b).} \label{fig2}
\end{figure}

In Fig.~\ref{fig2}, we compare simulations for the
susceptible-infected-susceptible (SIS) epidemic model \cite{SIS_ref1,SIS_ref2,SIS_ref3} with the corresponding predictions of a well-known MF theory \cite{PastorSatorras01,Barthelemy05}. In the MF theory, the fraction $i_k(t)$ of degree-$k$ nodes that are infected at time $t$ is given by the solution of the equation
\begin{equation}
	\frac{d i_k}{d t} = - i_k + \beta(1-i_k) k \Theta_k\,, \label{e1}
\end{equation}
where $\beta$ is the spreading rate, the recovery rate has been set to unity by choice of timescale, and
\begin{equation}
	\Theta_k = \sum_{\kp} P(\kp|k) i_{\kp} \label{e2}
\end{equation}
is the probability that any given neighbor of a degree-$k$ node is infected.  In Eq.~(\ref{e2}), $P(\kp|k)$ is the probability that an edge originating at a degree-$k$ node has a degree-$\kp$ node at its other end. Because degree-degree correlations are included, this version of the theory is called a \emph{correlated MF theory} (cMF).  A further simplification of the theory is possible if one assumes that the network is uncorrelated and is thus completely described by its degree
distribution $p_k$. In this case, which is termed \emph{uncorrelated MF} (uMF),
$P(\kp|k)$ in Eq.~(\ref{e2}) is replaced by $\kp p_{\kp}/z$ and
$\Theta_k$ becomes independent of $k$.  In Fig.~\ref{fig2}, we show predictions of both correlated and uncorrelated MF theories for the steady-state endemic infected fraction
\begin{equation*}
	I=\lim_{t \to \infty} \sum_k p_k i_k(t)
\end{equation*}	
for the AS Internet network \cite{asrel_url} and for the Electronic circuit (s838) network \cite{Milo04}.  The  MF theory is very accurate for the former, but it performs poorly for the latter.
 The result for the AS Internet network is particularly surprising in light of the fact  that the mean degree of this network is only 4. 
 The aforementioned naive argument that MF theory is accurate for high-degree nodes thus cannot
 account for the good performance of the theory in this low-$z$
 case, where 96\% of the nodes in the network have degree 10 or
 less (see Table~\ref{table1}).


\subsection{Voter Model}

\begin{figure}
\centering
\epsfig{figure=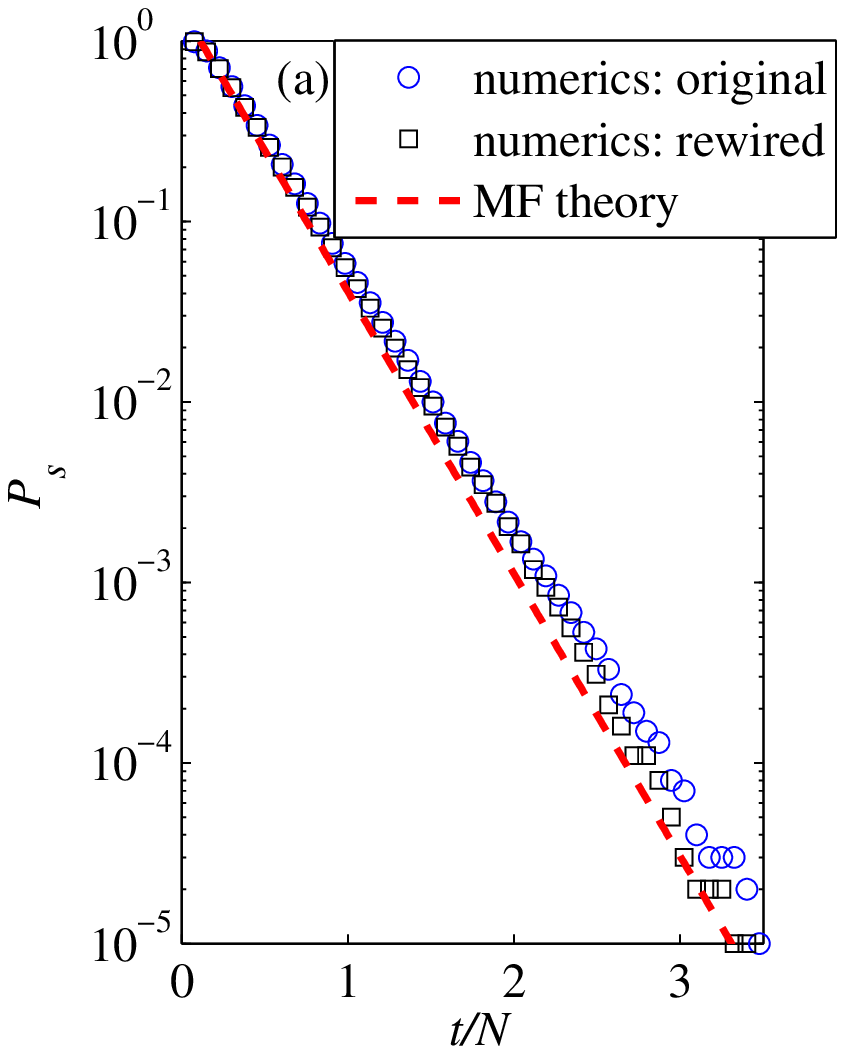,width=4cm}
\epsfig{figure=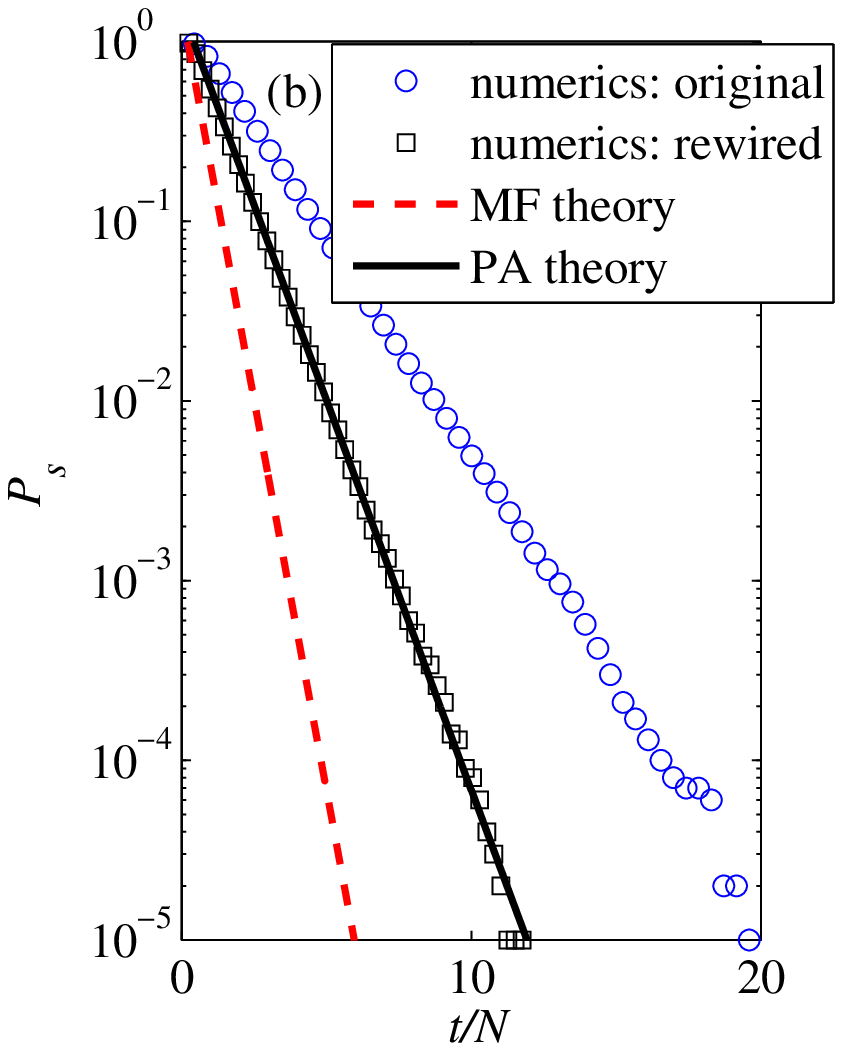,width=4cm}
 \caption{ (Color online) Survival probability for voter-model trajectories in the disordered state as a function of time on (a) \emph{C.~elegans} neural network \cite{Watts98} and (b) a synthetic clustered network generated as described briefly in the main text and in detail in Ref.~\cite{Gleeson09a} with $\gamma(3,3)=1$. Both networks contain approximately $N=300$ nodes.  The theory curves are from Refs.~\cite{Sood05,Vazquez08}.}
\label{fig3}
\end{figure}

As a third example, we consider the survival probability of disordered-state trajectories in the voter model \cite{Liggettbook} and compare it with the (uncorrelated) MF theory of Ref.~\cite{Sood05}. (For rigorous results for the voter model, see Refs.~\cite{Durrettbook,Durrett10} and references therein.)  At time $t=0$, each node is randomly assigned to one of two voter states.
In each time step (of size $dt=1/N$), a randomly-chosen node is updated by copying the state of one of its randomly-chosen neighbors.
 On finite networks, the dynamics eventually drive a connected component to complete order, in which all the nodes are in the same state. The survival probability $P_s(t)$ is defined as the fraction of realizations that remain in the disordered state at time $t$.
The red dashed curves in Fig.~\ref{fig3} gives the survival probability predicted by the MF theory of Ref.~\cite{Sood05}:
\begin{equation}\label{s1}
	P_s(t) \sim \frac{3}{2} \exp \left(-\frac{2 \left<k^2\right>}{z^2}\frac{t}{N} \right) \quad\text{ as $t\to\infty$}\,,
\end{equation}
and the black solid curve gives the results of the  pair approximation (PA) theory of Ref.~\cite{Vazquez08}:
\begin{equation}\label{s2}
	P_s(t) \sim \frac{3}{2} \exp \left(-\frac{2 (z-2)\left<k^2\right>}{(z-1)z^2}\frac{t}{N} \right) \quad\text{ as $t\to\infty$}\,.
\end{equation}

 In Fig.~\ref{fig3}(a), we show results for the \emph{C.~elegans} neural network \cite{Watts98}
 ($z \approx 14.46$), for which MF theory is very accurate. In
 Fig.~\ref{fig3}(b), we show results for a synthetic clustered network described in Ref.~\cite{Gleeson09a}: Every node in the network has degree 3 and is part of a single 3-clique.  Using the notation of Ref.~\cite{Gleeson09a}, this is called a $\gamma(3,3)=1$ network.  It can alternatively be described in the notation of Ref.~\cite{Newman09} as
having $p_{1,1}=1$, as every node is part of one triangle and has one single edge other than those belonging to the triangle. Clustering has a strong effect in this (low-$z$)
network; because of this clustering and dynamical correlations, MF
theory is very inaccurate. We can make the clustering negligibly small while keeping the degree distribution unchanged (at $p_k =
\delta_{k,3}$) by randomly rewiring the network to give a random
3-regular graph. However, even after this rewiring, the match to MF theory
is poor because dynamical correlations are still neglected.
The recent  PA theory for the voter model
\cite{Vazquez08} (see also \cite{Pugliese09}) accounts for the dynamical correlations and hence gives a good match to the survival probability on the rewired network but not on the clustered original network.

\setlength{\tabcolsep}{0.25cm}
\begin{turnpage}
\squeezetable
\begin{table}[h]
\begin{center}
\begin{ruledtabular}
\begin{tabular}{r|l|r|r|r|c|c|c|c|c|c|c|c|r}
\# & Network & $N$ & $z$ & $d$ & frac$(k \le 10)$ & $C$ & $\WT C$ & $r$ & $E_K$ & $E_S$ & $E_V$ & $E_V^{PA}$ & Ref(s)  \\ \hline
1 & Word adjacency: Spanish & 11558 & 7.45 & 942.58 & 0.91 & 0.38 & 0.02 & $-0.2819$ &
$-0.03$ & 0.01 & 0.21 & 0.04 & \cite{Milo04}\\
2 & Word adjacency: English & 7377 & 11.98 & 666.03 & 0.81 & 0.41 & 0.04 & $-0.2366$ & $-0.03$ & 0.01& 0.12 & 0.03  & \cite{Milo04} \\
3 & AS Internet & 28311 & 4.00 & 473.65 & 0.96 & 0.21 & 0.01 & $-0.2000$ & 0.08 & 0.03& 0.44 & 0.08  & \cite{asrel_url}\\
4 & Word adjacency: French & 8308 & 5.74 & 376.22 & 0.93 & 0.21 & 0.01 & $-0.2330$ & $-0.02$ & 0.03& 0.23 & 0.003  & \cite{Milo04}\\
5 & Marvel comics & 6449 & 52.17 & 338.16 & 0.25 & 0.78 & 0.19 & $-0.1647$  & $-0.03$ & 0.02 & 0.03 & 0.008 & \cite{Alberich02}\\
6 & Reuters 9/11 news & 13308 & 22.25 & 236.17 & 0.70 & 0.37 & 0.11 &
$-0.1090$ & $-0.01$ & 0.01 & 0.04 & -0.009& \cite{Johnson04}\\
7 & Word adjacency: Japanese & 2698 & 5.93 & 199.99 & 0.92 & 0.22 & 0.03 & $-0.2590$ & $-0.02$ & 0.02 & 0.24 & 0.01 & \cite{Milo04}\\
8 & Facebook Oklahoma & 17420 & 102.47 & 186.04 & 0.11 & 0.23 & 0.16 &
0.0737 & 0.03& 0.01& 0.02& 0.01  & \cite{Traud08,datadump}\\
9 & Corporate ownership (EVA) & 4475 & 2.08 & 113.85 & 0.98 & 0.01 &
0.00 & $-0.1851$ & 0.80& 0.06& 1.00 & 0.65  & \cite{Norlen02}\\
10 & Political blogs & 1222 & 27.36 & 100.07 & 0.45 & 0.32 & 0.23 &
$-0.2213$ & 0.04 & 0.01 & 0.05 & 0.009 & \cite{Adamic05}\\
11 & Facebook Caltech & 762 & 43.70 & 74.65 & 0.20 & 0.41 & 0.29 &
$-0.0662$ & 0.004 & 0.01& 0.04 & 0.009  & \cite{Traud08,datadump}\\
12 & Airports500 & 500 & 11.92 & 59.50 & 0.76 & 0.62 & 0.35 & $-0.2679$ & $-0.002$ & 0.06 & 0.29 & 0.19  & \cite{Colizza07, air500_url}\\
13 & \emph{C. Elegans} Metabolic & 453 & 8.94 & 51.57 & 0.86 & 0.65 &
0.12 & $-0.2258$ & 0.03 & 0.04& 0.18 & 0.04  & \cite{Duch05,cel_met_url}\\
14 & Interacting Proteins & 4713 & 6.30 & 32.92 & 0.84 & 0.09 & 0.06 &
$-0.1360$ & 0.14 & 0.04 & 0.19 & -0.03  & \cite{Colizza05,Colizza06,DIP_url}\\
15 & \emph{C. Elegans} Neural & 297 & 14.46 & 32.00 & 0.41 & 0.29 &
0.18 & $-0.1632$ & 0.10 & 0.03 & 0.09 & 0.006  & \cite{Watts98,cel_neur_url} \\
16 & Transcription yeast & 662 & 3.21 & 22.31 & 0.95 & 0.05 & 0.02 &
$-0.4098$ & 0.64 & 0.08& 0.70 & 0.36  & \cite{Milo02}\\
17 & Coauthorships & 39577 & 8.88 & 20.17 & 0.77 & 0.65 & 0.25 &
0.1863 & 0.22 & 0.08& 0.14 & 0.01  & \cite{Newman01b, cond_mat_url}\\
18 & Transcription \emph{E. coli} & 328 & 2.78 & 17.88 & 0.96 & 0.11 & 0.02 &
$-0.2648$ & 0.48 &0.09& 0.79 & 0.43  & \cite{Mangan03}\\
19 & PGP Network & 10680 & 4.55 & 13.46 & 0.90 & 0.27 & 0.38 & 0.2382 & 0.50 & 0.16& 0.45 & 0.17 & \cite{Guardiola02, Boguna04, PGP_url}\\
20 & Electronic Circuit (s838) & 512 & 3.20 & 4.01 & 1.00 & 0.05 &
0.05 & $-0.0300$ & 0.78 & 0.23& 0.62 & 0.23  & \cite{Milo04}\\
21 & Power Grid & 4941 & 2.67 & 3.97 & 0.99 & 0.08 & 0.10 & 0.0035 & 0.93 & 0.29& 0.90 & 0.65  & \cite{Watts98, power_url} \\
22 & $\g$-theory [$\g(3,3)=1$] & 1002 & 3   & 3 & 1.00  &
1/3     & 1/3          & N/A  & 0.91 & 0.89& 0.78 & 0.47  & \cite{Gleeson09a}\\
23 & $\g$-theory [$\g(3,3)=1$] & 10002 & 3   & 3 & 1.00  &
1/3     & 1/3          & N/A  & 0.97 & 0.88& 0.79 & 0.49   & \cite{Gleeson09a}\\
\end{tabular}
\end{ruledtabular}
\end{center}
\caption{Basic diagnostics and error measures for the networks used in this paper. All real-world data have been treated in the form of undirected, unweighted networks.  We only consider the largest connected component of each network, for which we enumerate the following
properties: total number of nodes $N$; mean degree $z$; mean first-neighbor degree $d$; fraction of nodes of degree 10 or less; clustering coefficients $C$ and $\WT C$ (defined, respectively, in
Eqs.~(3.6) and (3.4) of Ref.~\cite{Newman03a}); and Pearson degree correlation coefficient $r$.
The quantities $E_S$, $E_K$ and $E_V$ are the relative MF errors for the SIS,  Kuramoto, and voter models, $ E_V^{PA}$ is the relative PA error for the voter model. The last column gives the citation for the network in this paper's bibliography and/or the URL of a data file.
}
\label{table1}
\end{table}
\end{turnpage}


\begin{figure*}
\centering
\epsfig{figure=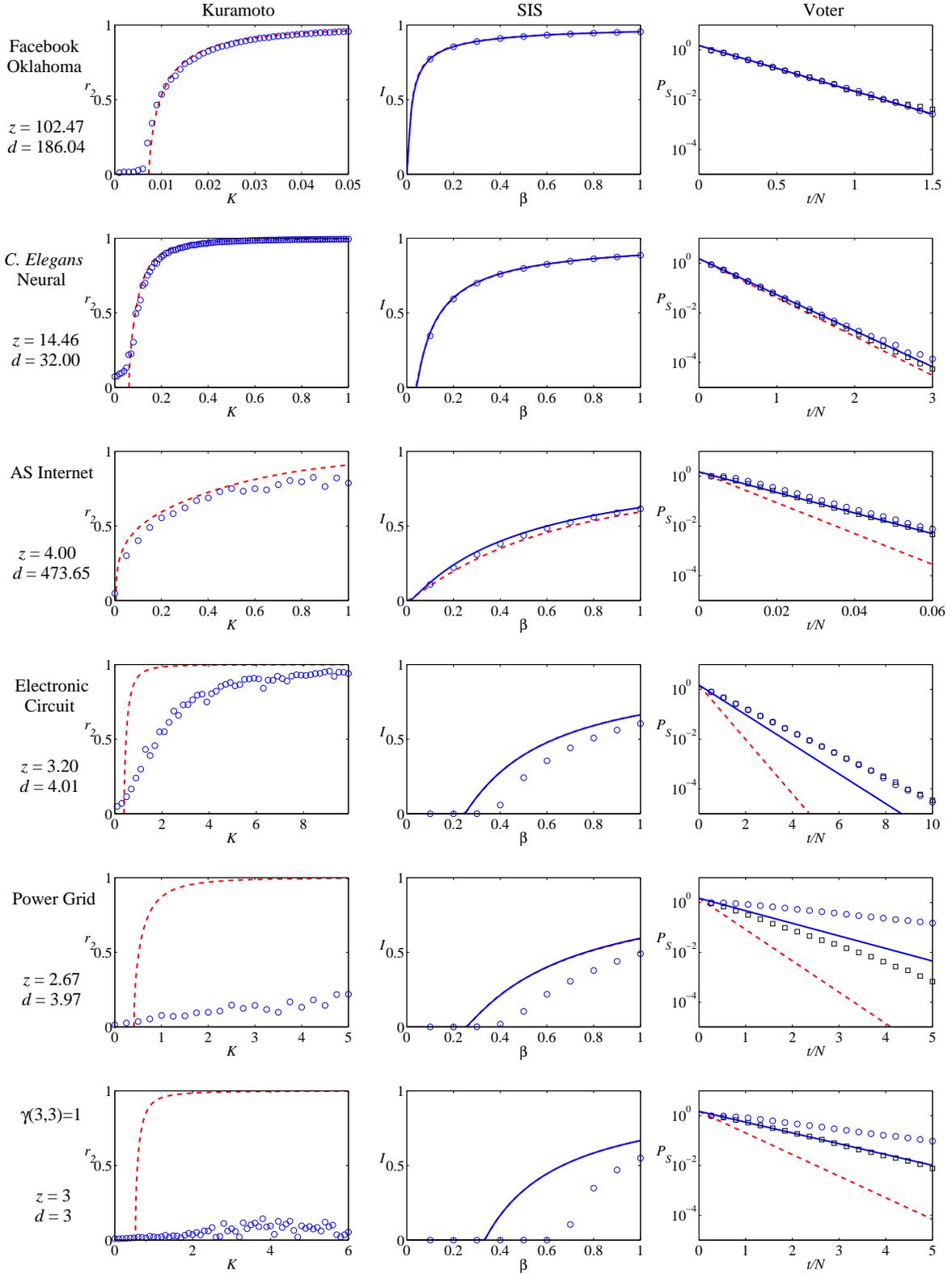,width=16cm}
 \caption{(Color online) Results for dynamics of Figs.~\ref{fig1}--\ref{fig3} for all 6 networks used in those figures.  Curves and symbols are  as in Figs.~\ref{fig1}--\ref{fig3}. For the voter model, black squares show the numerical results  obtained by rewiring networks in a manner that conserves both degree distribution and degree-degree correlations \cite{Melnik11}.
  } \label{fig4}
\end{figure*}


\section{Why is Mean-Field Theory Accurate?} \label{sec3}

Briefly summarizing our observations thus far, we have seen (i) situations in which high-$z$ networks exhibit good matches to MF theory, but also (ii)
some examples in which low-$z$ networks also have accurate MF
theories. For clarity, we have discussed only a few examples in detail, but these are representative of behavior observed for different dynamical processes on a variety of real-world networks.
 In Fig.~\ref{fig4}, we show additional examples for each of the three dynamical processes (the Kuramoto, SIS, and voter models) for each of the 6 networks used in Figs.~\ref{fig1}--\ref{fig3}.


\subsection{Mean-Field Accuracy for the SIS Model}

Clearly, the success of MF theories for dynamical systems on networks cannot be
explained purely in terms  of the mean degree $z$ of the underlying network. Figure~\ref{fig2}(a)
in particular gives an example in which MF theory works well on a
low-$z$ network.
To understand this seemingly surprising accuracy, we focus on the SIS model and consider how the state of a
node is updated as compared to the assumptions of MF theory.
Suppose the state of the degree-$k$ node $B_1$ is being
updated. In both the true dynamics and in MF theory, the updating
process depends on the state of the neighbors of $B_1$. Let's take node $B_2$ as a representative neighbor of $B_1$ and suppose that $B_2$ has degree
$\kp$. Under MF assumption (iii), the state of node $B_2$ is considered
to be independent of the state of node $B_1$.  This is why
Eq.~(\ref{e2}) involves the term
$i_\kp$, which is the probability that degree-$\kp$ nodes are in
the infected state, without any conditioning on the state of their
neighbor $B_1$ \footnote{Assumption (i) also appears in Eq.~(\ref{e2}) through the summation of $i_{k^\prime}$.  This supposes that the neighbors' states are independent of each other---i.e., that pairs of neighbors do not form part of a triangle with node $B_1$.}.

In reality, however, the states of nodes $B_1$ and $B_2$ exhibit dynamical correlations.  For example, during an epidemic, an infected node is more likely to have infected neighbors than a susceptible node. Such dynamical correlations can be included explicitly in pair-approximation theories \cite{Levin96,Eames02,House11}, and their neglect can be
a major source of error in MF theories. This suggests an important question: Under what
circumstances might the MF assumption of dynamical independence (iii) still give
accurate results for the update of node $B_1$? One can argue
that if the degree $\kp$ of node $B_2$ is sufficiently large, then the
state of node $B_2$ is influenced by many of its neighbors other than
node $B_1$, so the error in neglecting the particular dynamical
correlation between $B_2$ and $B_1$ is relatively small for the purpose of updating node $B_1$.
Conversely, if  the degree $\kp$ of node $B_2$ is small, then
node $B_1$ has a relatively strong influence on the state of node $B_2$,
and neglecting dynamical correlations between $B_2$ and $B_1$ when updating $B_1$ will yield
large errors. Hence, we expect MF theory to give reasonably accurate
updates for node $B_1$ if its neighbors have sufficiently high
degrees. Importantly, this argument relies only on the degree
$\kp$ of the nearest-neighbor nodes being high and gives no
restriction on the degree $k$ of the updating node itself.  (We remark that the use of networks with nearest-neighbor nodes of high degree has also been mentioned in studies of the Kuramoto model on networks \cite{Restrepo05,arenas08}.)

In short, we argue that MF theory gives relatively small error for
nodes with high-degree neighbors. More specifically, MF theory is
likely to be \emph{inaccurate} if many nodes do not have
high-degree neighbors.  This can happen, for example,
if low-degree nodes are connected preferentially to other low-degree
nodes (a sort of ``poor-club phenomenon'', akin to the rich-club phenomenon of high-degree nodes connecting preferentially to other high-degree nodes
\cite{Colizza06}). This suggests a simple but effective predictor of MF
accuracy: If the mean degree of first-neighbors $d=\sum_k p_k \sum_\kp \kp P(\kp|k)$ is
high, then MF theory can be expected
to be accurate.  As we show below, this rule of thumb works well for SIS dynamics on all of the networks that we have considered.

\begin{figure}
\centering
\epsfig{figure=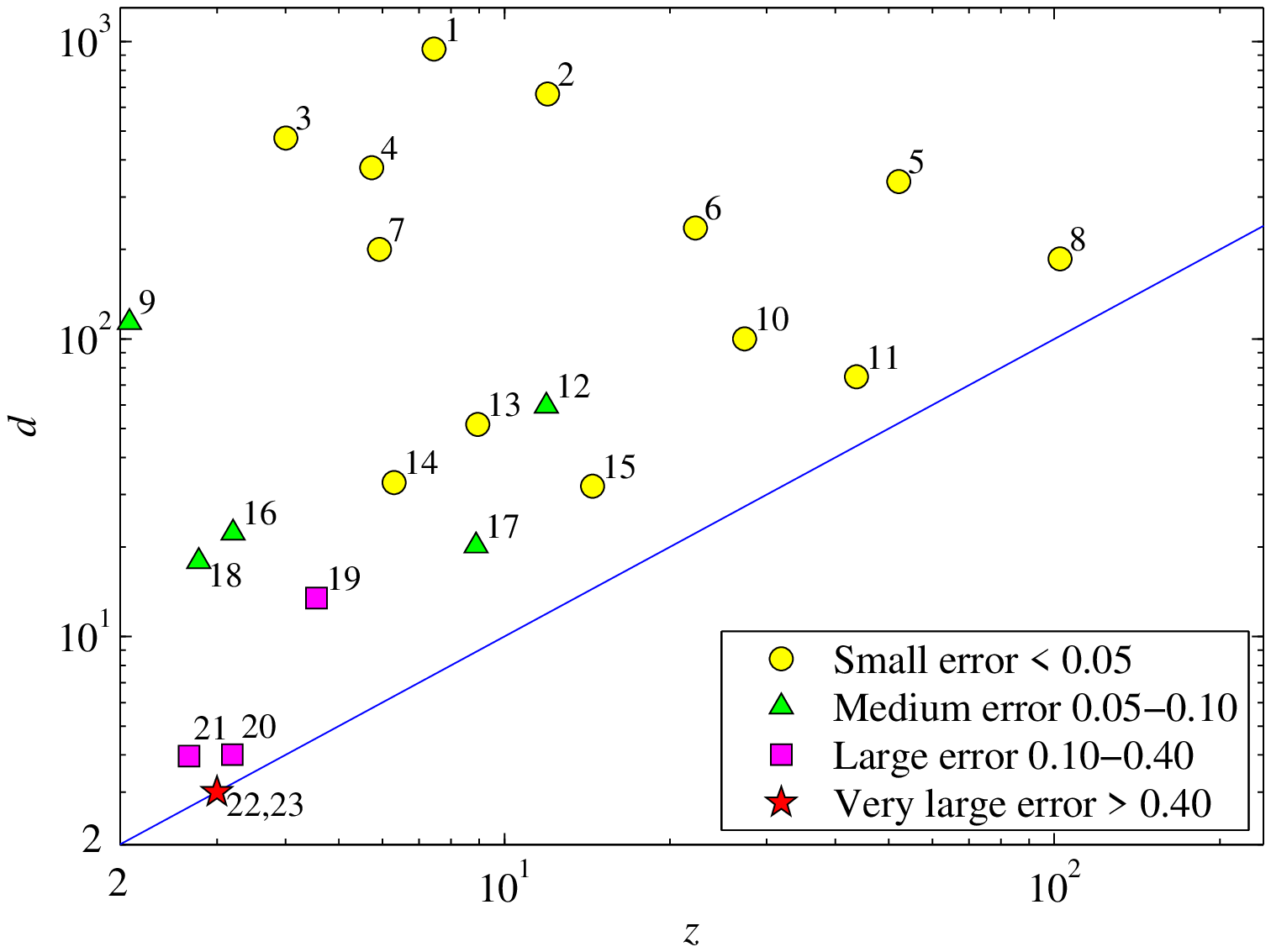,width=8.5cm}\caption{ (Color online) Values of mean degree $z$ and mean nearest-neighbor degree $d$ for many real-world networks. The numbers that label the networks are enumerated in column 1 of Table~\ref{table1}. The colors indicate the magnitude of the relative error $E_S$, defined in Eq.~(\ref{ES}), between cMF theory and numerical simulations for SIS dynamics with $I_{theory}=0.5$.} \label{fig5}
\end{figure}

In Fig.~\ref{fig5} we present results from
  numerical simulations of SIS dynamics and consider the final, steady-state fraction  $I$ of infected nodes (for each network we average
over an ensemble of more than several hundred realizations, in each of which 5\% of the
nodes are randomly chosen to be infected at $t=0$).
Because the quality of the MF approximation is known to
depend on the number of infected nodes \cite{Guerra10}, we further
compare errors for different networks by choosing the spreading rate
$\beta$ in each simulation so that the MF steady-state value
$I_\text{theory}$ equals 0.5.  Letting
 \begin{equation}
 	E_S = \frac{I_\text{theory}-I_\text {numerical}}{I_\text{theory}} \label{ES}
 \end{equation}
be the relative error between the theoretical predictions of
correlated MF theory (\ref{e1})--(\ref{e2}) and the numerical results,
we color each network's symbol in Fig.~\ref{fig5} by the value of
$E_S$ (see Table~\ref{table1} for the error values).
The results in Fig.~\ref{fig5} show clearly that
situations with low $d$ and low $z$ correspond to inaccurate cMF
predictions [see Figs.~\ref{fig1}(b), \ref{fig2}(b), and
  \ref{fig3}(b)] and that the high-$d$ situations (some of which also have small $z$) all have accurate cMF predictions, supporting our claim that
the fidelity of MF theory can be evaluated using $d$.

In Section~\ref{sec4}, we describe an alternative
  measure for predicting MF accuracy that uses inter-vertex distances
  \cite{Melnik11}. One can also construct other, more complicated
  measures (e.g., by computing the size of the connected cluster of low-degree nodes); however, the mean first-neighbor degree $d$ is
appealing because it is simple to calculate and understand, and it retains considerable explanatory power. We
note that similarly accurate results have been found for real-world
networks using MF theory for a discrete-time version of the SIS model
\cite{Gomez09,Guerra10}.

\subsection{Isolating the effect of dynamical correlations using synthetic networks}

\begin{figure}
\centering
\epsfig{figure=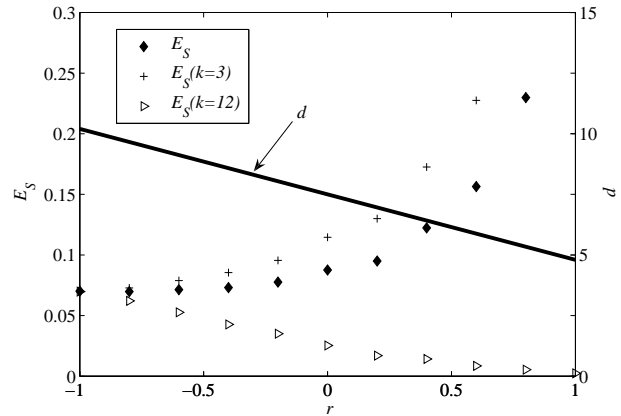,width=8.5cm}
 \caption{Error (diamonds) of cMF theory in predicting
   steady-state infected fraction for SIS dynamics as a function of
   Pearson correlation coefficient $r$ in the synthetic networks
   (described in the text) with $(k_1,k_2)=(3,12)$. Also shown are the
    errors for the class of degree-3 nodes (crosses) and for the
   class of degree-12 nodes (triangles): these are calculated similar to Eq.~(\ref{ES}), using $i_k(\infty)$ (from Eq.~(\ref{e1})) for $I_\text{theory}$, and the average infected fraction of degree-$k$ nodes in simulations for $I_\text{numerical}$.  The solid line gives the
   mean nearest-neighbor degree $d$.}\label{fig6}
\end{figure}
\begin{figure}
\centering
\epsfig{figure=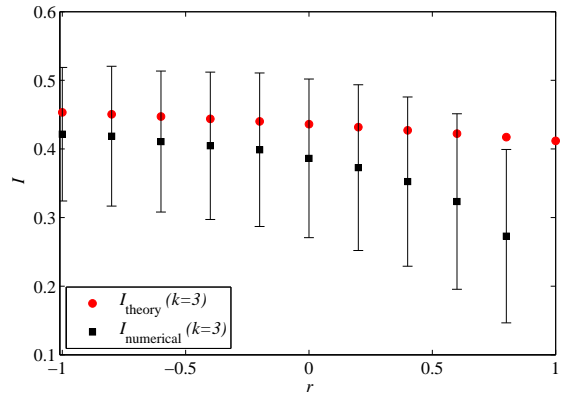,width=8.5cm}
 \caption{ (Color online) Mean (black squares) and standard deviation (error bars show one standard deviation above and below the mean) of the distribution of $f_i$ values (described in text) on the same networks as in Fig.~\ref{fig6}. The values of $I_\text{theory}$ are given by cMF theory, using $i_3(\infty)$ from Eq.~(\ref{e1}).}\label{fig7n}
\end{figure}
\begin{figure}
\centering
\epsfig{figure=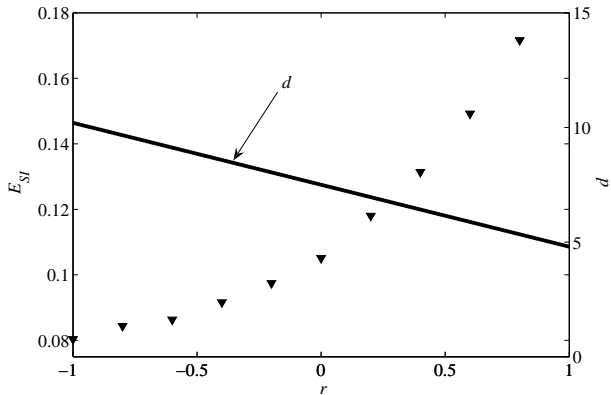,width=8.5cm}
 \caption{{ Relative errors $E_{SI}$ (symbols), given by Eq.~(\ref{ESI}), in the calculation of the fraction of SI edges using MF assumption (iii) on the same networks as in Fig.~\ref{fig6}. The solid line gives the
   mean nearest-neighbor degree $d$.}}\label{fig8n}
\end{figure}

We have argued above that the observed accuracy of MF theory on some real-world networks is due to their high $d$ values ameliorating the neglect of dynamical correlations [i.e., MF assumption (iii)]. However, real-world networks typically have high values for clustering coefficients and significant modular structures, so MF theory for such networks violates assumptions (i) and (ii) as well as assumption (iii). It might therefore be argued that the high-$d$ effect seen in Fig.~\ref{fig5} could be due to an improvement in the validity of assumptions (i) and/or (ii), which could in principle have a  larger impact than the high-$d$ improvement of assumption (iii). We investigate this  by now considering  SIS dynamics on synthetic random networks (with $N=10^4$ nodes) in which the transitivity and community structure are both negligible.  Thus, MF assumptions (i) and (ii) are both valid for these networks, and the error in MF theory can be due only to violations of assumption (iii).

  The first family of synthetic networks we use is described in Ref.~\cite{Dodds09}: Each node is either of low degree $k_1$ [with probability $p_{k_1}=k_2/(k_1+k_2)$] or of high degree $k_2$ [with probability $p_{k_2}=k_1/(k_1+k_2)$].  In order to create a network with a
prescribed degree-degree correlation coefficient $r$, we connect the nodes of each type preferentially to nodes of either the same or of opposite type.
We show in Fig.~\ref{fig6} how the aggregate error {$E_S$} (diamonds) depends on
the correlation coefficient $r$ for a specific case with $(k_1,k_2)=(3,12)$. Note that the mean degree $z$ of these networks is fixed ($z = 4.8$), but the mean first-neighbor degree $d$ decreases as $r$ increases. Figure \ref{fig6} illustrates that
the highest error for the MF theory occurs when $d$ is
lowest.  This is the fully assortative ($r=1$) case in which
low-degree nodes link only to other low-degree nodes, creating a low-$k$
connected cluster in which MF theory is inaccurate. At the other
extreme, the disassortative ($r=-1$) case has every low-degree node
linked only to high-degree nodes, with a consequent reduction in the
error of MF theory (and a high value of $d$). The trends of the errors in each degree class also support our argument: high-degree neighbors correspond to
lower MF error.

{ In Fig.~\ref{fig7n}, we show further details about the degree-3 nodes. To examine the importance of assumption (ii) we consider (for each value of $r$) an ensemble of $M=25$ realizations of the SIS process on identical copies of the synthetic network. At a fixed (large) time, we record the state---either infected or susceptible---of each node $i$. Taking the average over the $M$ realizations defines $f_i$, the fraction of realizations in which node $i$ is infected at the chosen time. We now consider  the distribution of $f_i$ values  over the set of nodes $i$ that all share the same degree $k$, noting that  assumption (ii) implies that these $f_i$ values should be identical for every node $i$ in a given $k$-class. Consequently, we plot in Fig.~\ref{fig7n}, for degree-3 nodes, the mean and standard deviation of the $f_i$ values for the same networks as used in Fig.~\ref{fig6}. Note the mean value equals $I_\text{numerical}(k=3)$, the average infected fraction of degree-3 nodes in simulations. This mean value deviates from the cMF theory prediction $i_3(\infty)$, giving the errors shown by the crosses in Fig.~\ref{fig6}. We note that the standard deviation of the $f_i$ values does not depend strongly on $r$ (and hence also does not depend strongly on $d$). Moreover, if all of the $f_i$ values for each $r$ value were to be replaced by their mean value---so that the standard deviation for $f_i$ was zero, in accord with  assumption (ii)---then the errors shown in Fig.~\ref{fig6} would be unaltered. In this sense, the violation of assumption (ii) is not the main source of the high-$d$ effect on the accuracy of MF theory.

To test for the existence of dynamical correlations---which are neglected under MF theory assumption (iii)---in the numerical results, we calculate the fraction $\phi_{SI}$ of SI edges (i.e., edges linking a susceptible node and an infected node) in the networks in steady state. We then compare this to the fraction of SI edges that would be obtained if no dynamical correlations were present. This is given by treating as independent the probabilities for nodes at each end of an edge to be in states S and I:
\begin{equation}
\phi_{SI}^{MF} = 2 \sum_{k,k^\prime} \rho_k \frac{k p_k}{z} P\left( k^\prime | k\right) (1-\rho_{k^\prime}), \label{phiMF}
\end{equation}
where $\rho_k$ is the fraction of infected degree-$k$ nodes in the network (hence $(1-\rho_\kp)$ is the fraction of degree-$\kp$ nodes that are susceptible). Note that we calculate $\rho_k$  from the numerical simulations---we do not use the corresponding MF theory values $i_k(\infty)$---so that we are directly testing the validity of assumption (iii) on the numerical data. The relative error of the MF assumption on the SI edge fraction can then be calculated in a similar fashion to Eq.~(\ref{ES}):
\begin{equation}
E_{SI} = \frac{\phi_{SI}^{MF}-\phi_{SI}}{\phi_{SI}^{MF}}. \label{ESI}
\end{equation}
If assumption (iii) were true, there would be no dynamical correlation effects and the fraction of SI edges could be computed directly from the fraction of infected nodes using (\ref{phiMF}), giving $E_{SI}=0$. However, Fig.~\ref{fig8n} shows that the error $E_{SI}$ increases as the mean first-neighbor distance decreases, which is similar to the trend of error $E_S$ in Fig.~\ref{fig6}. This evidence supports our claim that it is MF assumption (iii)---the neglect of dynamical correlations---that plays a dominant role in determining the accuracy of MF theory on high-$d$ networks.
}


The second family of synthetic networks is composed of networks with negligible degree-degree correlations, and is
generated using the configuration model \cite{Bollobas}. In these networks, $d = \left \langle k^2\right \rangle/z$, so the mean first-neighbor degree increases with the second moment of the degree distribution if the mean degree $z$ is fixed. For example, one can construct networks with $z=5$ with the degree probabilities
\begin{equation}
	p_3 = \frac{15}{17}(1-\alpha), \quad p_5 = \alpha, \quad p_{20}= \frac{2}{17}(1-\alpha)\,,
\end{equation}
and $p_k=0$ for all other $k$. Such a network has a fraction $\alpha$ of nodes with degree 5, and the remaining nodes have degrees 3 and 20. The mean first-neighbor degree for such networks is $d=11 - 6\alpha$, so it decreases linearly with $\alpha$.  By comparing numerical simulations for the SIS model with MF theory (not shown), we find that the error magnitude  $|E_S|$ increases monotonically with $\alpha$.  It takes the value 0.07 at $\alpha=0$ (with $d=11$) and the value 0.16 at $\alpha=1$ (with $d=5$). Similar to the correlated synthetic networks of Fig.~\ref{fig6}, the higher values of $d$ thus correspond to lower values of the error.

{The evidence from both families of synthetic networks suggests} that the high-$d$ effect that we have observed can be due only to its impact on dynamical correlations [i.e., assumption (iii)].  In real-world networks, $d$ can presumably affect the validity of all three MF assumptions.  Further work is required to understand which assumption(s) have the strongest impact on MF accuracy in such situations.


\subsection{Mean-Field Accuracy for Other Dynamical Processes}

\begin{figure}
\centering
\epsfig{figure=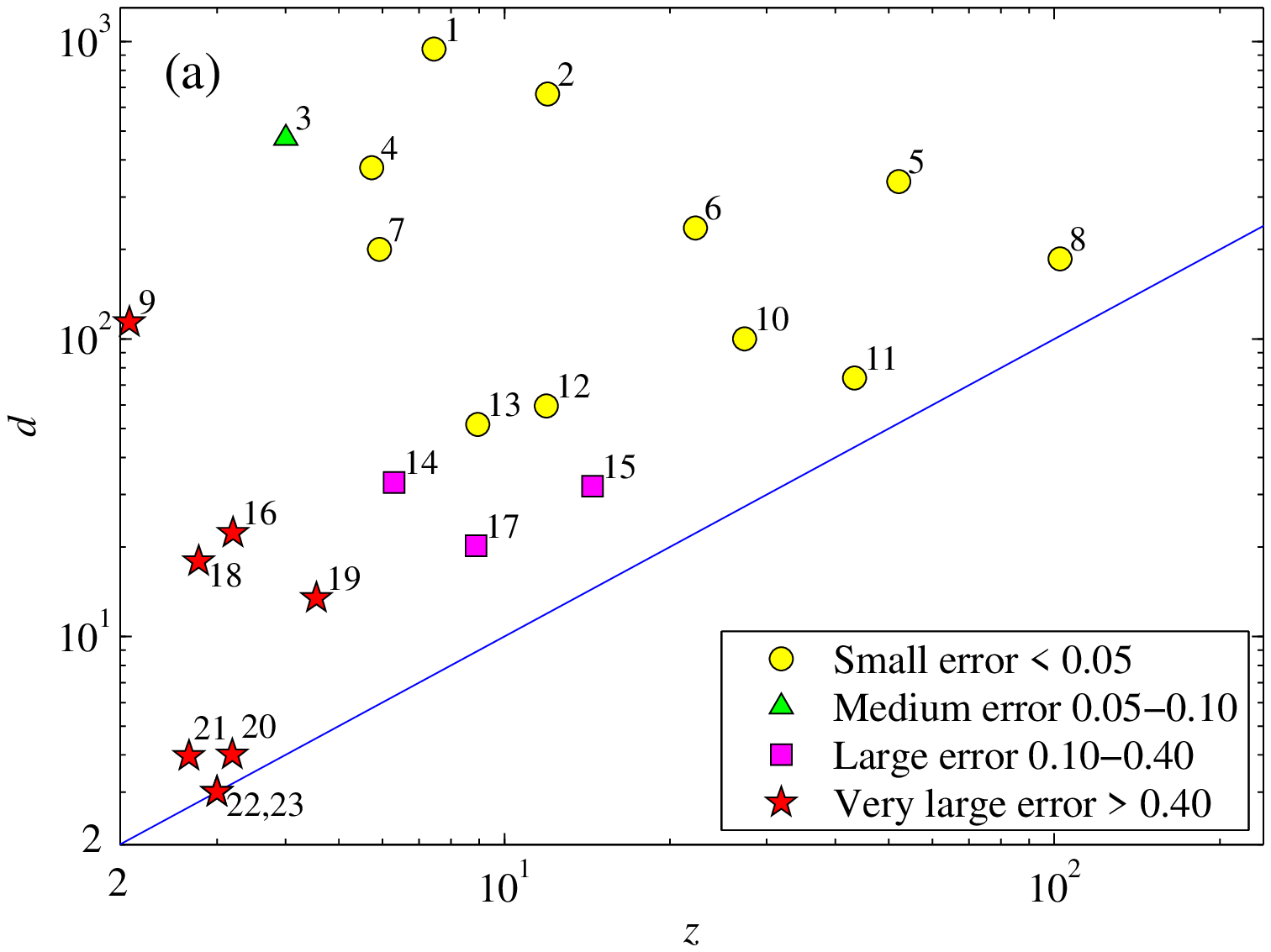,width=8.5cm}
\epsfig{figure=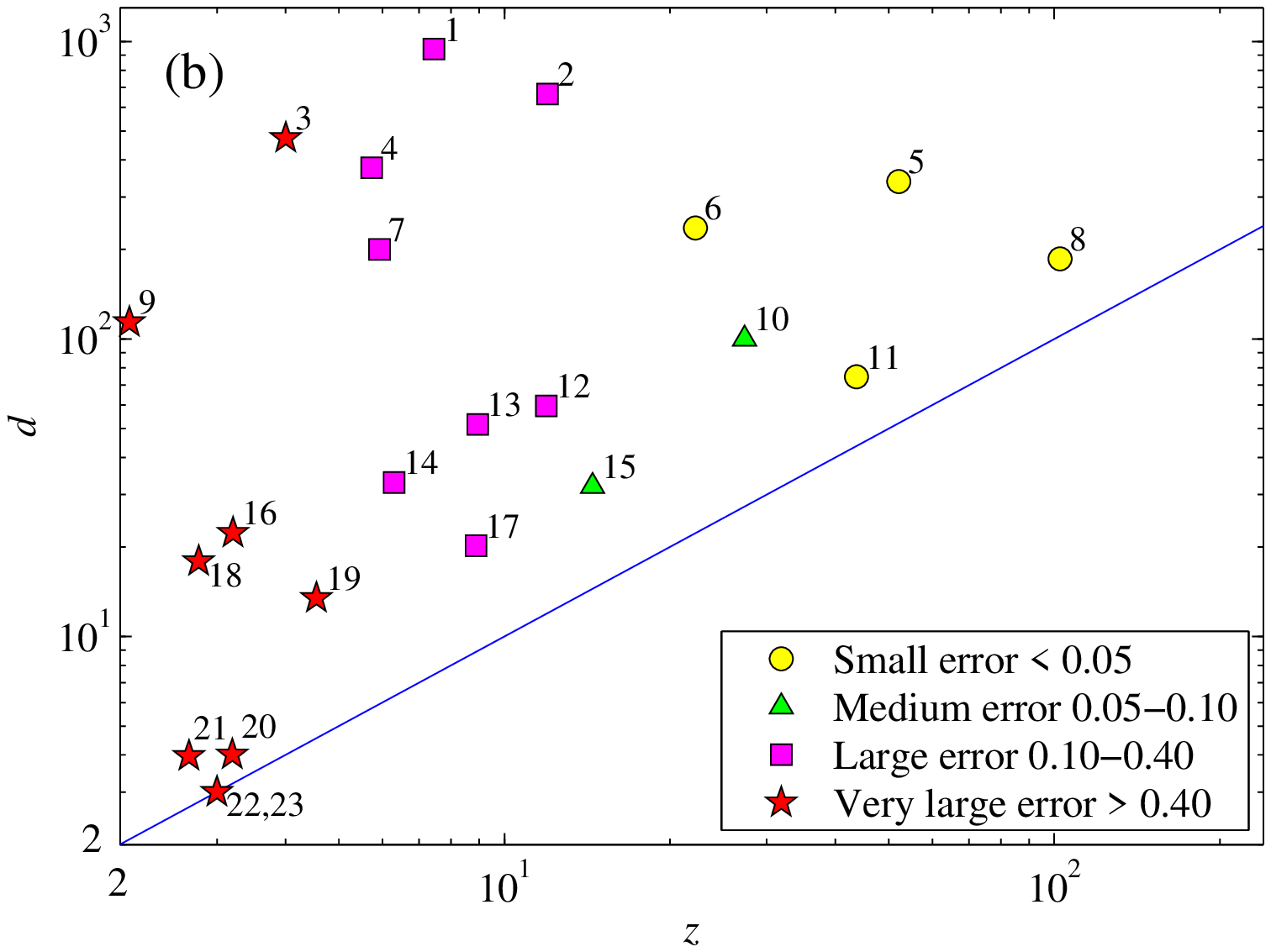,width=8.5cm}
\epsfig{figure=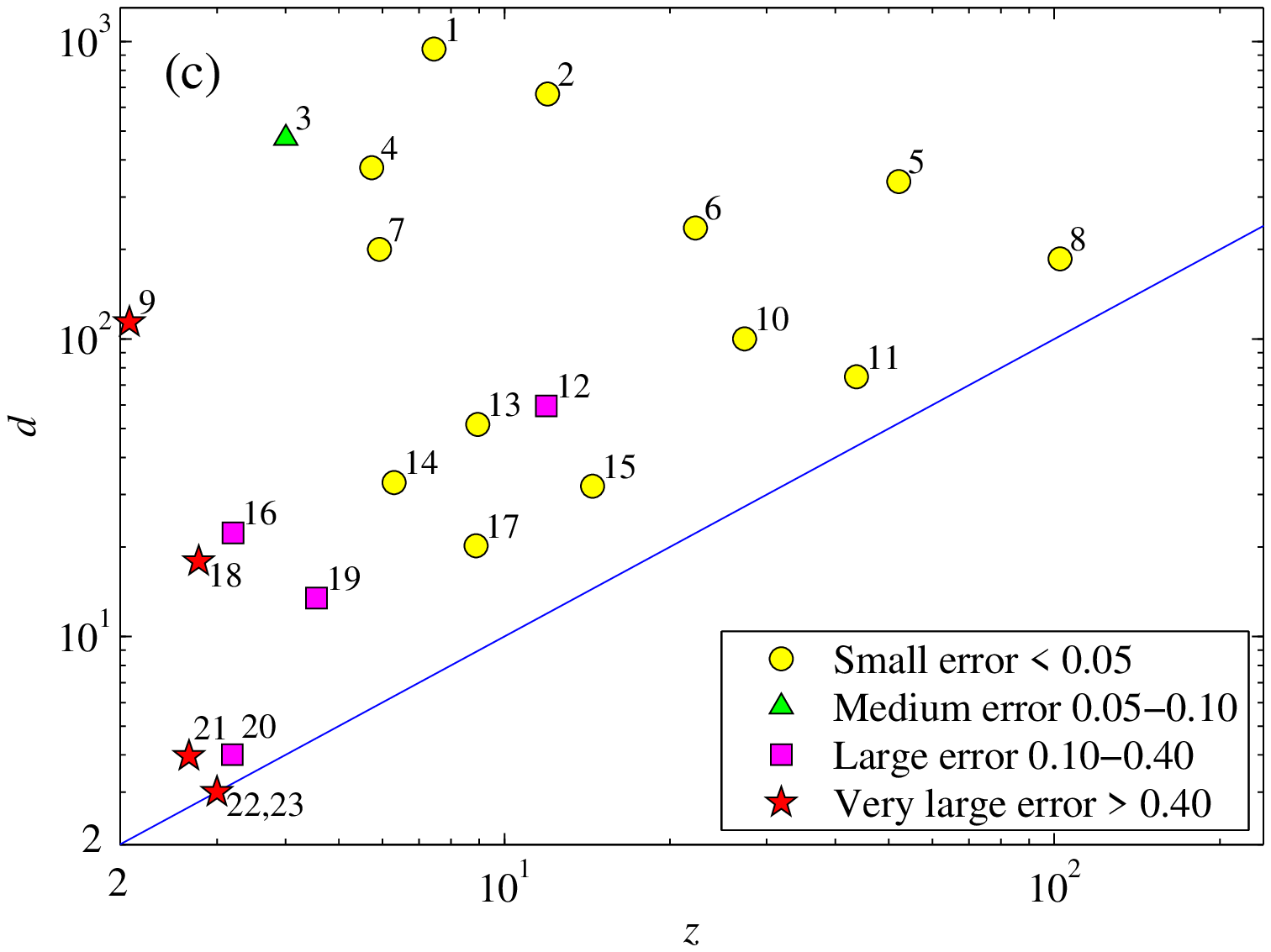,width=8.5cm}
 \caption{ (Color online) As in Fig.~\ref{fig5}, except that color indicates the magnitude of the relative error, which is determined by comparing numerical simulation results to: (a) MF theory for the Kuramoto order parameter $r_2$ (with $r_{2\text{ theory}}=0.6$), from Eq.~(\ref{EK}); (b) MF theory for the voter model survival probability $P_s(t)$ at the time $t$ (with $P_{s\text{ theory}}=10^{-2}$) from Eq.~(\ref{EV}); and (c) PA theory for the voter model survival probability. } \label{fig7}
\end{figure}

The argument that we have given above for the usefulness of $d$ as a measure of MF accuracy is specific to SIS dynamics, where the quantity of interest is the (ensemble-averaged) infected fraction of nodes. We now consider error measures on the $(z,d)$ plane for the other dynamics studied here, the Kuramoto and voter models. For the Kuramoto model, we define an error measure in terms of the $r_2$ order parameter from Eq.~(\ref{r2eqn}) as
 \begin{equation}
	E_K = \frac{r_{2\text{ theory}}-r_{2\text{ numerical}}}{r_{2\text{ theory}}}\,, \label{EK}
\end{equation}
which we evaluate at the value of $K$ for which $r_{2\text{ theory}}=0.6$. Similarly, a measure of relative error for the voter model is given by
\begin{equation}
	E_V=\frac{\log_{10}\left(P_{s\text{ theory}}\right) - \log_{10}\left(P_{s\text{ numerical}}\right)}{\log_{10}\left(P_{s\text{ theory}}\right)}\,, \label{EV}
\end{equation}
where $P_{s\text{ theory}}$ and $P_{s\text{ numerical}}$ are the survival probabilities given by Eq.~(\ref{s1}) and by numerical simulations, respectively.  We evaluate these quantities at the time $t$ that corresponds to a survival probability of one percent (in MF theory): $P_{s\text{ theory}}(t)=10^{-2}$. This definition reflects the vertical difference between the dashed curve and the symbols in Fig.~\ref{fig3} at a specific value of $t$.

We give the measured values for $E_V$ and $E_K$ for all real-world
networks in Table~\ref{table1}, and Figures~\ref{fig7}(a) and
\ref{fig7}(b) show how these values depend on the mean degree $z$ and
mean first-neighbor degree $d$ of the networks. The Kuramoto model
exhibits a similar pattern to SIS (compare Fig.~\ref{fig7}(a) to
Fig.~\ref{fig5}): high-$d$ networks have lower errors than low-$d$
networks.  However, the high-$d$ effect does not seem to impact the
voter model [see Fig.~\ref{fig7}(b)] in the same way.  We can
understand this by contrasting the predicted quantities for SIS and
for the voter model.  For SIS, the error is low when MF theory
accurately predicts the fraction $I(t)$ of infected nodes.  In the
voter model, the quantity corresponding to $I(t)$ is the fraction of
nodes in one of the two voter states, which we denote by
{$I_V(t)$}. When initial states are randomly
assigned, MF theory  predicts that $I_V(t)$
is conserved, which implies that $I_V(t)=1/2$ for
all $t$. Numerical simulations on real-world networks also give
$I_V(t)=1/2$, but only when one averages over an
ensemble of realizations. In any single realization on a network of
finite size, fluctuations eventually lead to the entire network
becoming ordered---i.e., all nodes eventually share the same
opinion. (In half of the realizations, this shared opinion is one voter state; in the other half it is the other state.)  It is this
single-realization ordering process that is measured by the survival
probability $P_s(t)$. In this respect, the voter model is different
from both the SIS and Kuramoto models, in which the trajectory of a
typical realization is qualitatively similar to an ensemble-average of
all trajectories.

In order to better capture quantities of higher-order than
{$I_V(t)$}, such as $P_s(t)$, it is necessary to
approximate the dynamical correlations between nodes using, for
example, a pair-approximation method. In Fig.~\ref{fig7}(c), we show
the magnitude of the PA error for the voter model; we still measure
the error using Eq.~(\ref{EV}), but we use the PA theory of
  Eq.~(\ref{s2}) instead of the MF theory of Eq.~(\ref{s1}). Observe
  the improvement in accuracy over the MF theory of
  Fig.~\ref{fig7}(b), particularly for high-$d$ networks. Similar to
  MF for SIS (see Fig.~\ref{fig5}) and for Kuramoto [see
    Fig.~\ref{fig7}(a)], only networks with both low $z$ and low $d$
  have relative errors that significantly exceed 5\%.


\section{Comparison with an alternative measure}
\label{sec4}

 Using numerous examples of real-world and synthetic networks, we have illustrated that the mean degree of first-neighbors $d$ is a good indicator of the accuracy of MF theories for a variety of dynamical processes on networks. One can also construct more complicated accuracy measures, which may in general depend on the dynamics under scrutiny. In Ref.~\cite{Melnik11}, for example, we examined (among other dynamics) the accuracy of the bond percolation theory of~\cite{Vazquez03} by comparing its predictions with numerical calculations of the sizes of the largest connected component for several real-world networks.  We showed that a good measure of the error is given by the quantity \begin{equation}
	q = \frac{\ell-\ell_1}{z}\,, \label{qeqn} \end{equation} where $\ell$ is the mean inter-vertex distance in the original
(clustered) network, $\ell_1$ is the corresponding mean distance in a rewired version of the network (using a rewiring process that preserves degree-degree correlations but reduces clustering), and $z$ is the mean degree. Noting that the bond percolation theory of Ref.~\cite{Vazquez03} is of pair-approximation (PA) type, in contrast to the mean field theories on which we focus in this paper, it is nevertheless of interest to examine the relation between $q$ and the mean first-neighbor degree $d$ that we have identified in this paper as an indicator of MF theory accuracy for several dynamical processes.

\begin{figure}
\centering
\epsfig{figure=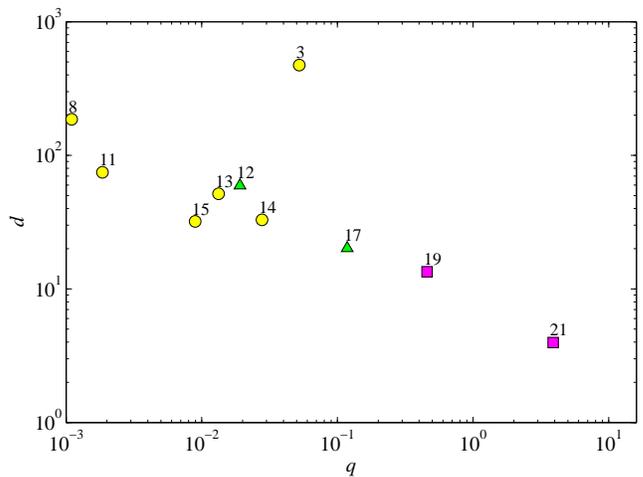,width=8.5cm}
 \caption{(Color online) Location of real-world networks in the
   $(q,d)$ parameter plane, where $q$ is the error measure
   (\ref{qeqn}), which was shown in Ref.~\cite{Melnik11} to be
   correlated with the error of PA theory for bond percolation.  We
   use the same symbols as in
   Fig.~\ref{fig5}.} \label{fig8}
\end{figure}
In the $(q,d)$ parameter plane of Fig.~\ref{fig8}, we show the positions of those real-world networks from Fig.~\ref{fig5} which were also examined in Ref.~\cite{Melnik11}. The expected relationship between $q$ and $d$ is revealed: Networks with high $d$ have low $q$ (and hence, according to Ref.~\cite{Melnik11}, have low error for bond percolation PA theory), while low $d$ values correspond to high $q$ and hence to large errors. Thus, despite its simplicity, $d$ performs well when compared to other more involved measures. As noted above, there is scope for further work on developing more complicated diagnostics (such as $q$) for predicting MF accuracy for various dynamics, but the use of the $d$ value is appealing because it is simple to calculate and aids in understanding the underlying causes of MF inaccuracies.

\section{Conclusions}
\label{sec5}
In summary, we have shown that MF theory works best for networks in which low-degree nodes (if
present) are connected to high-degree nodes (i.e., for
networks that are either disassortative by degree or have high mean degree $z$).
Remarkably, it is not necessary that the mean degree of the network be large for MF theory to work well---at least for ensemble-averaged quantities [see Figs.~\ref{fig5} and \ref{fig7}(a)]. In addition to the 21 real-world networks that we have studied, we have presented evidence from synthetic networks to support our hypothesis.
We stress that our error measures focus on behavior far from critical points; the accuracy of MF for the calculation of phase transition points (such as the value of $K$ for the onset of synchronization in the Kuramoto model \cite{Hong02, Gomez-Gardenes07} or the SIS epidemic threshold \cite{Durrett10,Castellano10,Parshani10,Gleeson11}) is a topic for future work.

Based on our results for the voter model in Sec.~\ref{sec3}, we expect that similar conclusions should hold for the applicability of pair-approximation theories (such as those in refs.~\cite{Eames02,Pugliese09}) for dynamics on
real-world networks. Although PA theories account for the dynamical
correlations that plague MF theories, they remain vulnerable to
the effects of network clustering and modularity when $d$ is
low.  For example, Fig.~\ref{fig3}(b) gives an example in which PA theory works
well only on the rewired (and hence unclustered) version of a (low-$d$)
network. This suggests that PA theory (like MF theory) is most accurate for real-world networks
with either high mean degree or high mean first-neighbor degree.


\section*{Acknowledgements}

We acknowledge funding from Science Foundation Ireland (06/IN.1/I366, MACSI 06/MI/005, and 09/SRC/E1780) (JPG and JAW), from INSPIRE: IRCSET-Marie Curie
International Mobility Fellowship in Science Engineering and
Technology (SM), from the James S. McDonnell Foundation (\#220020177)
(MAP), from the EPSRC for support of MOLTEN (EP/I016058/1) (JAW) and
from the NSF (DMS-0645369) (PJM).  We thank Adam D'Angelo and Facebook
for providing the Facebook data used in this study, J.-P. Onnela and
Dan Fenn for supplying us with the data that they assembled for  Ref.~\cite{Onnela10}, and Alex
Arenas, Mark Newman, CAIDA, and Cx-Nets Collaboratory for making
publicly available other data sets that we used in this paper.  We
thank Andrea Baronchelli, Claudio Castellano, Rick Durrett, and two
anonymous referees for helpful comments.








\end{document}